\documentclass[twocolumn,journal]{IEEEtran}
\usepackage[T1]{fontenc}
\usepackage[latin9]{inputenc}
\usepackage{color}
\usepackage{array}
\usepackage{varwidth}
\usepackage{amsmath}
\usepackage{amssymb}
\usepackage{graphicx}
\usepackage[bookmarks=true,bookmarksnumbered=true,bookmarksopen=true,bookmarksopenlevel=1,
 breaklinks=false,pdfborder={0 0 0},pdfborderstyle={},backref=false,colorlinks=false]
 {hyperref}
\hypersetup{pdftitle={Your Title},
 pdfauthor={Your Name},
 pdfpagelayout=OneColumn, pdfnewwindow=true, pdfstartview=XYZ, plainpages=false}

\makeatletter

\providecommand{\tabularnewline}{\\}
\newenvironment{cellvarwidth}[1][t]
    {\begin{varwidth}[#1]{\linewidth}}
    {\@finalstrut\@arstrutbox\end{varwidth}}

\let\oldforeign@language\foreign@language
\DeclareRobustCommand{\foreign@language}[1]{%
  \lowercase{\oldforeign@language{#1}}}

\usepackage[caption=false,font=footnotesize]{subfig}

\usepackage{colortbl}
\usepackage{cite}

\usepackage{graphicx}
\usepackage{braket}

\makeatother

\begin{document}
\title{ On the SNR Statistics in Coupled-Core Multi-Core Fiber Transmissions
with Mode-Dependent Loss}
\author{Chiara~Lasagni,\IEEEmembership{~Member,~IEEE,} Paolo~Serena,~\IEEEmembership{Senior~Member,~IEEE,}
Alberto~Bononi,~\IEEEmembership{Senior~Member,~IEEE}, \\
Lucas A. Zischler,\IEEEmembership{~Student~Member,~IEEE,} Giammarco~Di~Sciullo,\IEEEmembership{~Student~Member,~IEEE,}
Antonio~Mecozzi,~\IEEEmembership{Fellow,~IEEE,~Fellow,~Optica,}
and Cristian~Antonelli,~\IEEEmembership{Senior~Member,~IEEE,~Fellow,~Optica}\thanks{Manuscript received {*}{*}{*}{*}{*}{*}{*} {*}{*}, 2026. This work
has been supported by University of Parma through the action \emph{Bando
di Ateneo 2023 per la ricerca, }the Next Generation EU under the Italian
National Recovery and Resilience Plan (NRRP), Mission 4, Component
2, Investment 1.3, CUP B53C22003970001, partnership on \textquotedblleft Telecommunications
of the Future\textquotedblright{} (PE00000001 - program \textquotedblleft RESTART\textquotedblright ),
and by the European Union\textquoteright s Grant Agreement 101120422
- Quantum Enhanced Optical Communication Network Security (QuNEST).\protect \\
This work has been submitted to the IEEE for possible publication.
Copyright may be transferred without notice, after which this version
may no longer be accessible.}\thanks{C. Lasagni, P. Serena, and A. Bononi are with the Department of Engineering
and Architecture, Universit\`a degli Studi di Parma, Parma, 43124,
Italy, and with National Inter-University Consortium of Telecommunications
- CNIT, Italy (e-mail: \protect\href{http://chiara.lasagni@unipr.it}{chiara.lasagni@unipr.it},
\protect\href{http://paolo.serena@unipr.it}{paolo.serena@unipr.it};
\protect\href{http://alberto.bononi@unipr.it}{alberto.bononi@unipr.it}).}\thanks{L. A. Zischler, G. Di Sciullo, A. Mecozzi, C. Antonelli are with the
Department of Physical and Chemical Sciences, Universit\`a degli
Studi dell'Aquila, L\textquoteright Aquila, 67100, Italy, and with
National Inter-University Consortium of Telecommunications - CNIT,
Italy (e-mail: \protect\href{http://giammarco.disciullo@graduate.univaq.it}{giammarco.disciullo@graduate.univaq.it},
\protect\href{http://lucas.zischler@univaq.it}{lucas.zischler@univaq.it},
\protect\href{http://antonio.mecozzi@univaq.it}{antonio.mecozzi@univaq.it},
\protect\href{http://cristian.antonelli@univaq.it}{cristian.antonelli@univaq.it}). }\thanks{Color versions of one or more of the figures in this paper are available
online at http://xxxxxxxxxx.xxxx.xxx.}\thanks{Digital Object Identifier xx.xxxx/JLT.xxxx.xxxxxxx}}
\markboth{Journal of Lightwave Technology}{Chiara Lasagni \MakeLowercase{\emph{et al.}}: Title}
\IEEEpubid{}
\maketitle
\begin{abstract}
We investigate the impact of mode-dependent loss (MDL) on the statistics
of the signal-to-noise ratio (SNR) in coupled-core multi-core fiber
(CC-MCF) systems. Through numerical simulations, we present an in-depth
analysis of the impact of MDL on received amplified spontaneous emission
(ASE) noise and nonlinear interference (NLI), as well as their joint
contribution to the SNR. We show that MDL induces different statistics
on the two noises and discuss the differences with single-mode polarization-dependent
loss. Moreover, we investigate the impact of spatial mode dispersion
(SMD) on the MDL-induced impairment, offering insights on their joint
effects on ASE and NLI.
\end{abstract}


\IEEEpeerreviewmaketitle{}

\section{Introduction}

\IEEEPARstart{O}{\emph{}ptical} communication systems implementing
space-division multiplexing (SDM) have the benefit of leveraging multiple
light paths to increase capacity compared to standard single-mode
fiber (SMF). Multi-core and multi-mode fibers, or combinations of
the two, are common SDM options\cite{Puttnam21}. However, depending
on the fiber design, the spatial paths may experience significant
crosstalk, causing the signals propagating in each spatial path to
become evenly mixed \cite{Ho_Kahn_book,Antonelli_springer}. This
coupling regime occurs in multi-core fibers with narrow core separation,
known as coupled-core multi-core fibers (CC-MCFs), and within mode
groups of multi-mode fibers.

In the case of CC-MCFs, a multiple-input multiple-output (MIMO) equalizer
is required to remove linear crosstalk and untangle the signals. Nevertheless,
the strong coupling among the cores is beneficial in reducing the
accumulation of impairments during propagation, such as spatial mode
dispersion (SMD), mode-dependent loss (MDL), and nonlinear interference
(NLI) \cite{Antonelli_tutorial}. The slower accumulation rate of
SMD \cite{Hayashi_MCF}, $\text{ps/\ensuremath{\sqrt{\text{km}}}}$
rather than ps/km, is of particular relevance to reduce the complexity
of the MIMO in terms of a reduced number of taps \cite{Antonelli_stokes}.
 As a result, CC-MCFs have consistently shown the potential of achieving
ultra-high data rates \cite{Ryf:19,BeppuOECC21,Rademacher_JLT21,Rademacher_JLT22,Kawai25},
also in deployed conditions \cite{LuisECOC25}, ultimately reaching
the order of petabits per second \cite{Luis25}.

The reduction of the accumulated MDL is an important factor at system
level, as MDL is a major source of impairment limiting system capacity
\cite{Ho2011,Antonelli2015,Beppu2022}. Mode-dependent loss refers
to the different random losses/gains experienced by the polarizations
and spatial modes propagating in the optical system \cite{Ho_Kahn_book},
and it generalizes the concept of polarization-dependent loss (PDL)
in single mode fibers to the multidimensional scenario of SDM. Sources
of MDL span from optical fibers to multiplexers, amplifiers, and other
optical devices \cite{Arikawa2022,Mazur_ECOC21,Mazur_20,Rademacher_NatureComm21},
rendering essential a characterization of the link MDL for effective
system design and operation \cite{Ospina24,Choutagunta2018}. Contrary
to mode dispersion, MDL cannot be completely removed by digital signal
processing (DSP) \cite{Shtaif_PDL08}. For instance, even with zero-forcing
equalization with perfect channel state information, MDL remains on
the detected noises. In fact, as both amplified spontaneous emission
(ASE) noise introduced by the amplifiers and the fiber NLI arise distributed
along the link, they accumulate a different amount of MDL compared
to the signal. As a consequence, MDL degrades the signal-to-noise
ratio (SNR) and makes it a random variable inducing outage events
on the system performance.

For these reasons, evaluating the impact of MDL on the system performance
is crucial, and it has been the focus of several studies \cite{Antonelli_stokes,Ho2011,Antonelli2015,Mello2020,Arikawa_JLT25,Lucas_JLT25,Andrusier2014,Winzer2011,Lucas_MMSE_JLT25,Gholamipourfard_JLT25}.
However, in these studies, the NLI has been either neglected or treated
as unaffected by MDL. Nevertheless, similar single-mode studies highlighted
significant differences between the impact of PDL on the linear and
nonlinear noise \cite{Rossi14,Rossi2019,Serena20PDL}.

In this work, which extends our preliminary study in \cite{Lasagni_ECOC24},
we investigate the impact of MDL and SMD on the SNR in the presence
of both ASE noise and NLI. Exploiting numerical split-step Fourier
method (SSFM) simulations, we study the MDL-induced statistics of
the linear and nonlinear noise individually, to highlight the differences
between the two noises. We first carry out the simulations in the
absence of SMD, to focus solely on the effect of MDL. Then, we consider
a more comprehensive scenario where both MDL and SMD impair the system,
and we estimate the statistics of the SNR in different conditions.

The numerical approach adopted in this work offers significant insights
into the impact of MDL on system performance, which are often hidden
in the experiments. Furthermore, it allows to extend the analysis
to the nonlinear regime in SDM links, which remains largely unexplored
experimentally. However, the SSFM's numerical effort is significant,
as it requires many runs to adequately sample the SNR statistics.
For this reason, analytical or semi-analytical models are extremely
useful in reducing the computation times in practical scenarios. In
this paper, we also adapt the PDL-GN model, which was first published
in \cite{Serena20PDL} to study the impact of PDL on the NLI in SMF,
to the case of CC-MCFs in a simplified scenario without SMD. This
extension proves itself accurate against numerical simulations.

The paper is structured as follows. Section II provides an overview
of MDL in optical systems based on CC-MCFs, offering comparisons with
single-mode PDL. Section III focuses on the impact of MDL on the statistics
of the SNR, with and without mode dispersion. Finally, Section IV
presents our conclusions.

\section{MDL in CC-MCF systems\protect\label{sec:MDLmatrix}}

The accumulation of MDL in an MCF-based link can be described through
a waveplate model by generalizing the single-mode theory \cite{Damask}
to $2N$ dimensions \cite{Ho_Kahn_book}, with $N$ accounting for
the number of cores and the factor $2$ for polarization degeneracy.
We divide the link into the concatenation of $K$ independent sections.
A section describes the MDL accumulated in a segment of the optical
fiber or within an optical device, like an amplifier or a switch.
Mode-dependent loss couples the $2N$ polarizations through a matrix
$\mathbf{M}^{(k)}$, whose singular value decomposition (SVD) can
be expressed as:
\begin{align}
\mathbf{M}^{(k)}=e^{\frac{1}{2}\overline{g}^{(k)}}\mathbf{V}^{(k)}\boldsymbol{\Lambda}^{(k)}\mathbf{U}^{(k)\dagger}, & \quad k=1,\ldots,K\label{eq:Mk}
\end{align}
where $\mathbf{V}^{(k)}$ and $\mathbf{U}^{(k)}$ are $2N\times2N$
random unitary matrices representing random mode coupling at the beginning
and end of the section, and the dagger denotes Hermitian conjugate.
The uncoupled propagation of the field in each section is described
by the diagonal matrix $\boldsymbol{\Lambda}^{(k)}=\mathrm{diag}[e^{\frac{1}{2}g_{1}^{(k)}},\ldots,e^{\frac{1}{2}g_{2N}^{(k)}}]$,
where the gains satisfy $\sum_{i}g_{i}^{(k)}\!=\!0$, as we factored
out the average MDL $\overline{g}^{(k)}$ in (\ref{eq:Mk}). 

The matrix $\mathbf{M}$ modeling the system MDL is given by the
concatenation of the $K$ section matrices, namely:
\begin{align}
\mathbf{M}=\mathbf{M}^{(K)}\cdots\mathbf{M}^{(2)}\mathbf{M}^{(1)}=\mathbf{V}\boldsymbol{\Lambda}\mathbf{U}^{\dagger}\label{eq:totalM}
\end{align}
with $\boldsymbol{\Lambda}=\mathrm{diag}[e^{\frac{1}{2}g_{1}},\ldots,e^{\frac{1}{2}g_{2N}}]$,
where we define the system MDL by the factors $g_{i}$, $i=1,\ldots,2N,$
which are the logarithm of the squared singular values of the total
transfer matrix $\mathbf{M}$, or, equivalently, the logarithm of
the eigenvalues of the MDL operator $\mathbf{M}\mathbf{M^{\dagger}}=\mathbf{V}\left(\boldsymbol{\Lambda}\right)^{2}\mathbf{V}^{\dagger}$
\cite{Ho2011,Andrusier2014}. 

Equation (\ref{eq:Mk}) can be generalized to include mode dispersion.
In this case, in the Fourier domain, the generic $i$th element on
the main diagonal of the matrix $\boldsymbol{\Lambda}^{(k)}$ becomes
$e^{\frac{1}{2}g_{i}^{(k)}-j\omega\tau_{i}^{(k)}}$, with $\tau_{i}^{(k)}$
the time delay experienced by the $i$-the space and polarization
mode in section $k$, and $\omega$ the angular frequency shift. 
In the case of lumped MDL, the total matrix in (\ref{eq:totalM})
is a concatenation of frequency-independent MDL matrices and MDL-free
frequency-dependent fiber sections introducing mode dispersion.

Due to the random nature of each section, the system MDL impairing
the polarizations is a random variable, whose distribution has been
studied and approximated analytically both in SMF \cite{Karlsson_PDL}
and MCF \cite{Ho2011} transmissions.

\begin{figure}[t]
\begin{centering}
\includegraphics[width=0.75\linewidth]{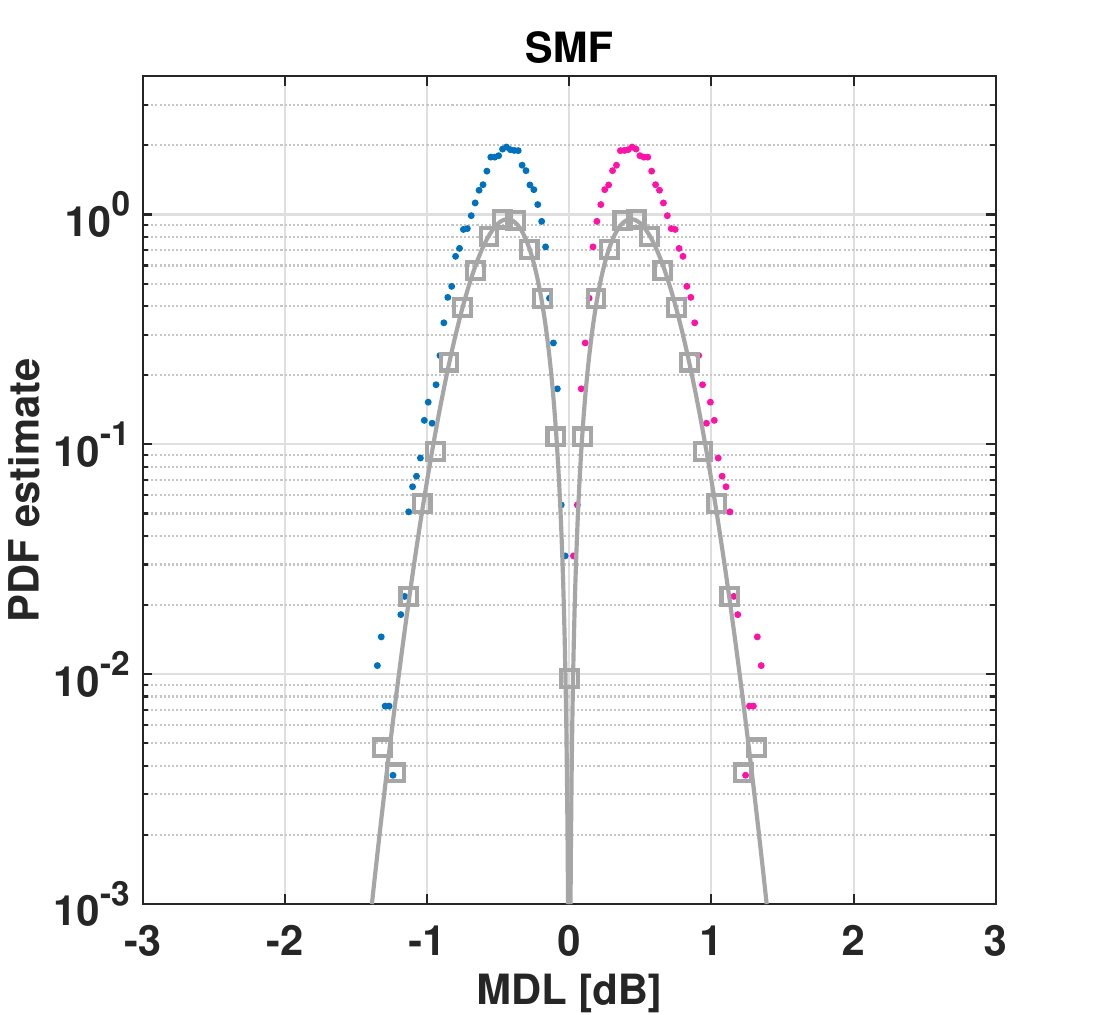}\\
(a)\\
\includegraphics[width=0.75\linewidth]{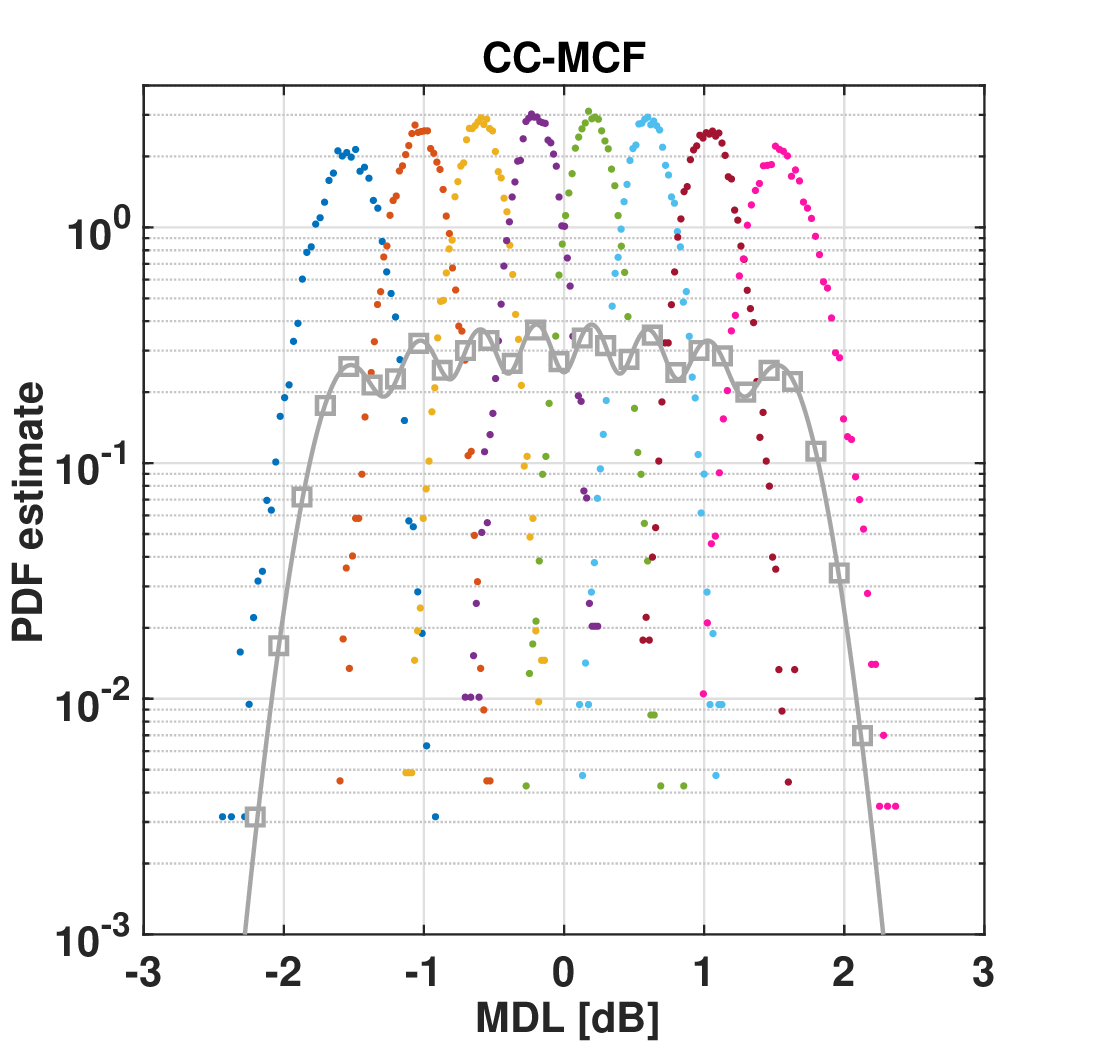}\\
(b)
\par\end{centering}
\caption{Probability density function of the system MDL expressed in dB. Results
obtained from $10^{4}$ random realizations of the MDL transfer matrix,
built by concatenation of $K=256$ sections. Top: single core. Bottom:
four strongly coupled cores. Squares: distribution of a randomly-picked
singular value of $\mathbf{M}$. Dots: marginal distribution of the
$2N$ order statistics. Solid lines: expressions from \cite{Ho2011}.
\protect\label{fig:MDL_histograms}\textbf{\textcolor{red}{{} }}}
\end{figure}

To better understand the differences between the single-core and multi-core
scenarios, we implemented the previously described waveplate model
to numerically estimate the probability density function (PDF) of
the MDL in the two cases. We considered a transmission along $K\!=\!256$
sections. Similar to \cite{Ho_Kahn_book}, we used the same matrix
$\boldsymbol{\Lambda}^{(k)}$ in each section, with $g_{i}^{(k)}=(-1)^{i}\sigma_{g}$.
The value of $\sigma_{g}$ was set to $0.014$ in the four-core, and
adapted to $0.0076$ in the single-core case, in order to investigate
both cases at the same MDL-induced reduction in average spectral efficiency
per mode \cite{Andrusier2014}. We collected $10^{4}$ random realizations
of the MDL transfer matrix, and, for each realization, we estimated
the system MDL from its singular values. Figure \ref{fig:MDL_histograms}
shows with square markers the PDF of a randomly selected singular
value of $\mathbf{M}$, i.e., the mixture distribution. The dotted
PDFs refer to the $2N$ singular-values order statistics, i.e., the
smallest (leftmost) up to the largest (rightmost) singular values.
 The top figure refers to a single core while the bottom to four
strongly-coupled cores.

In the single-core case, the two marginal PDFs in Fig. \ref{fig:MDL_histograms}(a)
refer to the smallest and largest singular value, which, depending
on the specific realization, can impair either the $x$ or $y$ polarization.
Therefore, as one polarization gains power, the other is attenuated,
resulting in a clear difference between the best and worst polarization.
Figure \ref{fig:MDL_histograms}(a) also shows that the numerical
PDF of the MDL agrees with the analytical two-sided Maxwellian distribution
(solid line) \cite{Karlsson_PDL}. 

In the multi-core case, due to the strong coupling among space and
polarization modes, the PDF of the MDL exhibits a different shape
compared to the Maxwellian distribution observed in the single-core
case. Namely, the mixture distribution has a multimodal distribution
with $2N\!=\!8$ peaks, in agreement with the analytical expression
reported in \cite{Ho2011} (solid line), and the experimental measurements
shown in \cite{Cappelletti_JLT24}. It is worth noting that the dotted
PDFs refer to the singular-value order statistics and not to the individual
polarizations in the cores. As a consequence, the two polarizations
of a given core may not have symmetric PDFs as in the SMF case, and
they may both experience a gain or a loss.

Despite the useful insights provided by the MDL PDF, at a system
level the MDL is often quantified in terms of standard deviation $\sigma_{\mathrm{MDL}}$
of the generic element $g_{i}$, $i=1,\ldots,2N$, or mean peak-to-peak
MDL \cite{Choutagunta2018,Mello2020}. The latter metric is defined
as the mean of the $10\mathrm{log_{10}(\cdot)}$ of the ratio between
the maximum and minimum squared singular values of the system matrix
$\mathbf{M}$ \cite{Ho2011}:
\begin{align}
\left\langle \mathrm{MDL}_{\mathrm{dB}}\right\rangle = & \left\langle 10\mathrm{log_{10}}\left(e^{(g_{\mathrm{max}}-g_{\mathrm{min}})}\right)\right\rangle \label{eq:meanMDL}\\
= & 4.343\left\langle \left(g_{\mathrm{max}}-g_{\mathrm{min}}\right)\right\rangle \,\nonumber 
\end{align}
with $\left\langle \cdot\right\rangle $ denoting ensemble averaging. 

\section{Impact of MDL on the signal-to-noise ratio}

\begin{figure}[t]
\begin{centering}
\includegraphics[width=0.99\linewidth]{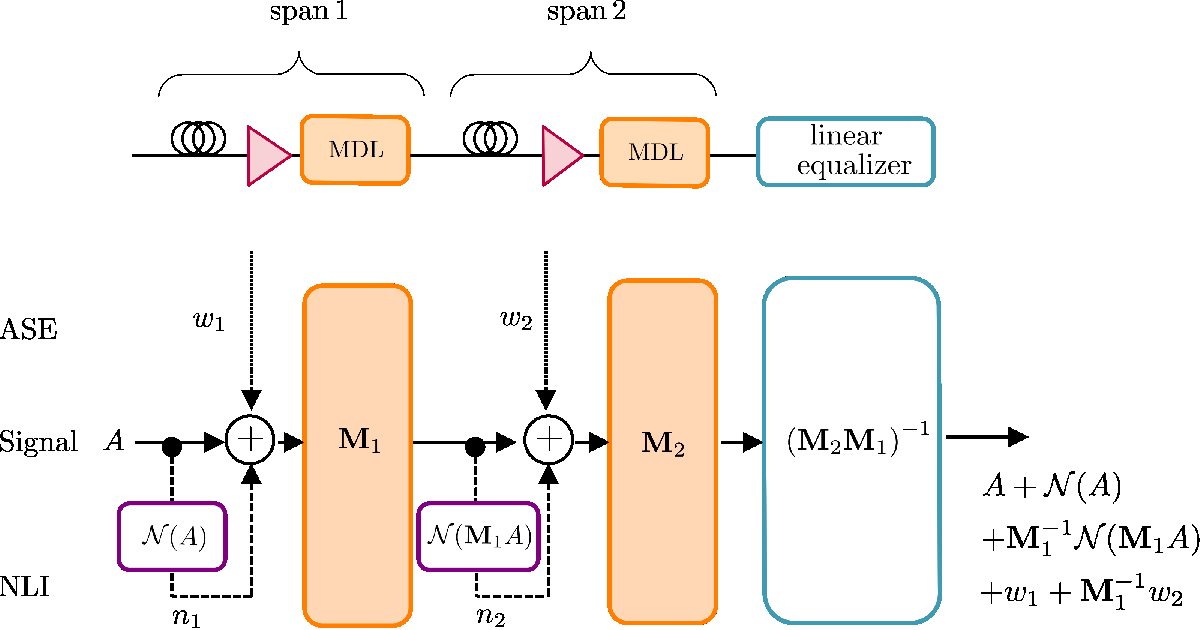}
\par\end{centering}
\caption{Schematic MDL accumulation in a two-span optical system with lumped
MDL and zero-forcing linear equalization. $A$: signal. $w_{i}$ and
$n_{i}$: ASE noise and perturbative NLI, respectively, arising in
the $i$th span. \protect\label{fig:sketch-1}}
\end{figure}

Assuming ASE and NLI as independent additive noises, the SNR of a
generic frequency channel in a given core can be conveniently expressed
as:
\begin{equation}
\frac{1}{\mathrm{\text{SNR}}}=\frac{N_{x}+N_{y}}{P_{x}+P_{y}}=\frac{1}{\text{\ensuremath{\text{SNR}_{\text{ASE}}}}}+\frac{1}{\mathrm{\text{SNR}_{\text{NLI}}}}\label{eq:SNR}
\end{equation}
where $N_{x,y}$ and $P_{x,y}$ represent the noise and signal power
in the $x$ and $y$ polarization. $\mathrm{\text{\ensuremath{\text{SNR}_{\text{ASE}}}}}$
and $\mathrm{\text{SNR}_{\text{NLI}}}$ are the linear and nonlinear
SNR accounting only for ASE or NLI as a source of noise, respectively.
 The presence of MDL changes the ASE and NLI power as they accumulate
MDL differently compared to the signal. This concept is sketched in
Fig. \ref{fig:sketch-1}, for the simple case of a two-span system
with lumped MDL at each span end. The figure highlights that, contrary
to the ASE noise $w_{i}$, the NLI $n_{i}$ arising in the $i$th
span depends on the signal $A$ and, thus, on the MDL at previous
coordinates. Such dependency is a key factor that distinguishes the
two types of noise. It is represented in the figure through the nonlinear
operator ${\cal N}(A)$ which, under perturbative assumptions, can
be reasonably approximated by a multilinear (cubic) map of the individual
signal frequencies entering the span \cite{SerenaSDMGN,Dar16}. 

To investigate the different impact of the residual MDL on ASE and
NLI after signal equalization, we focused on three scenarios: i) only
ASE, ii) only NLI, iii) ASE + NLI. We considered the transmission
of independent wavelength-division multiplexed (WDM) signals in each
core of a $10\times100$ km optical link based on CC-MCFs with $N\!=\!4$
cores and lumped MDL after each span. Fiber propagation was implemented
using SSFM simulations solving the Manakov equation for CC-MCFs \cite{Antonelli_tutorial}.
The fiber under test had core-independent attenuation $0.2$ dB/km,
chromatic dispersion $17\,\text{ps/(nm\ensuremath{\cdot\text{km)}}}$,
and nonlinear coefficient $1.267/N\!=\!0.3167$ $\text{1/(W\ensuremath{\cdot\text{km)}}}$.
The fiber SMD coefficient varied depending on the system configuration.
The amplifiers had frequency flat gain of $20$ dB and noise figure
equal to $6$ dB. 

\begin{figure}[t]
\begin{centering}
\includegraphics[width=1\linewidth]{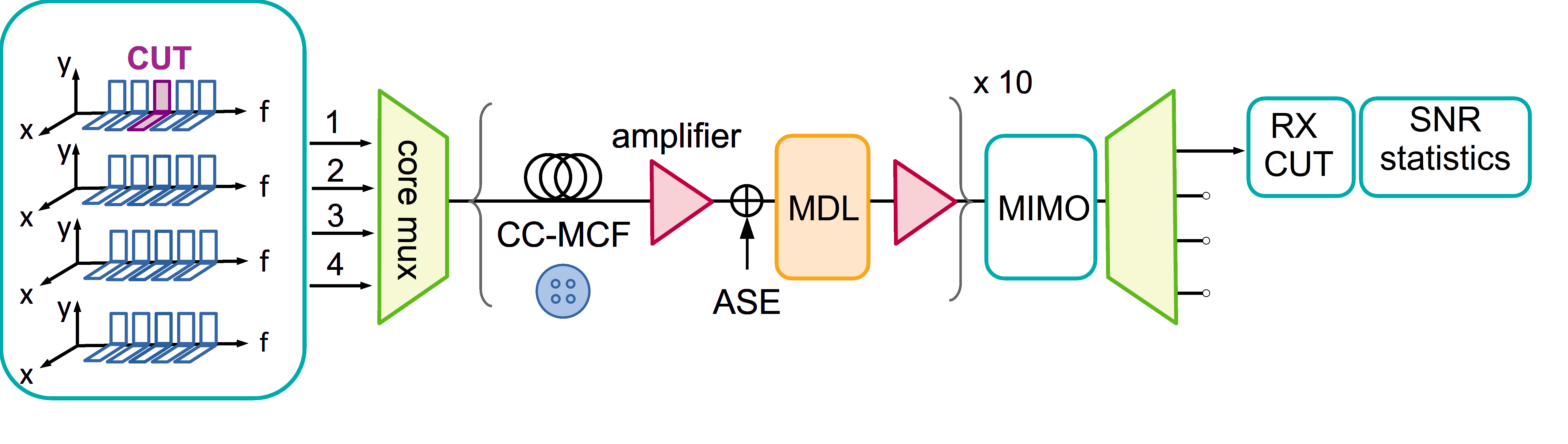}
\par\end{centering}
\caption{Sketch of the transmission under test. Five dual-polarizarion WDM
channels transmitted in each core of 100-km four-core CC-MCFs. The
MDL is applied at span end, with the last amplifier recovering the
mode-averaged loss. After ten identical spans, ideal equalization
of the linear effects and coherent detection are performed, and the
SNR of the channel under test is estimated for each MDL realization.\protect\label{fig:sketch}}
\end{figure}

To emulate the MDL introduced by an amplifier, we placed an MDL element
at each span end, as shown in Fig. \ref{fig:sketch}. The MDL element
was modeled by cascading a frequency-independent matrix, as described
in Sec. \ref{sec:MDLmatrix}, and a noiseless amplifier equalizing
the mode-averaged loss. Each MDL element had peak-to-peak MDL of
$1$ dB ($\sigma_{g}^{2}\!=\!0.015$), yielding a link peak-to-peak
MDL of $4.8$ dB according to (\ref{eq:meanMDL}). 

We considered the transmission of a WDM signal composed of five dual-polarization
channels per core, with a symbol rate of $64$ Gbaud and channel spacing
of $75$ GHz. We transmitted sequences of 65536 complex Gaussian distributed
symbols, well above the number required by the maximum channel walk-off,
to improve the accuracy of Monte Carlo estimations. The dual-polarization
power $P_{x}+P_{y}$ of each WDM channel power was set to 5 dBm, corresponding
to the optimal value maximizing the SNR in the absence of MDL and
SMD.

At the receiver, we optically compensated all linear effects accumulated
during propagation by inverting the channel transfer matrix. After
the detection of the dual-polarization channel under test (CUT) and
ideal average carrier phase recovery, we estimated the CUT SNR from
the received constellation. As depicted in Fig. \ref{fig:sketch},
the CUT was the central WDM channel of core 1. The statistical equivalency
of the remaining cores is verified in the next section.

Mode dispersion, when present, was implemented with the waveplate
model of the optical fiber described in Sec. \ref{sec:MDLmatrix}
with $100$-m long waveplates. For each configuration of link parameters,
the step was updated with the constant local error criterion \cite{Zhang08}.
The initial step was halved until ten seeds of the detected SNR reached
saturation within $0.05$ dB after the first $100$-km span.

In the absence of SMD, we also adopted the semi-analytical Gaussian
noise (GN) model in \cite{Serena20PDL} to estimate the PDF of the
SNR. Although this model was originally proposed and validated for
single-mode transmissions in the presence of PDL, thanks to the strong
coupling among the cores, it can be applied as well in our scenario
by adapting it to a higher dimension of $2N$ space and polarization
modes.  

\subsection{SNR statistics without mode dispersion \protect\label{subsec:SNR-statistics-without}}

\begin{figure}
\begin{centering}
\includegraphics[width=0.75\linewidth]{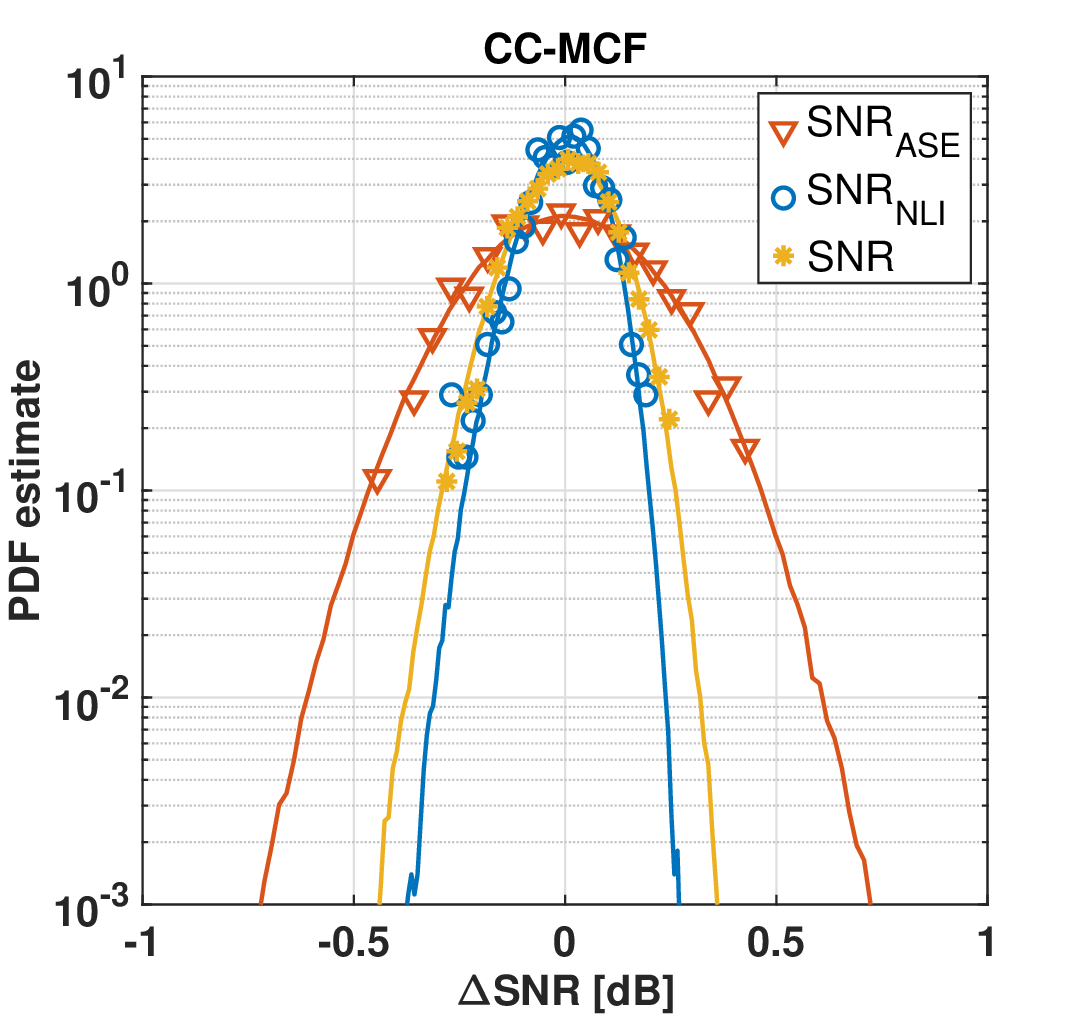}\\
(a)\\
\includegraphics[width=0.75\linewidth]{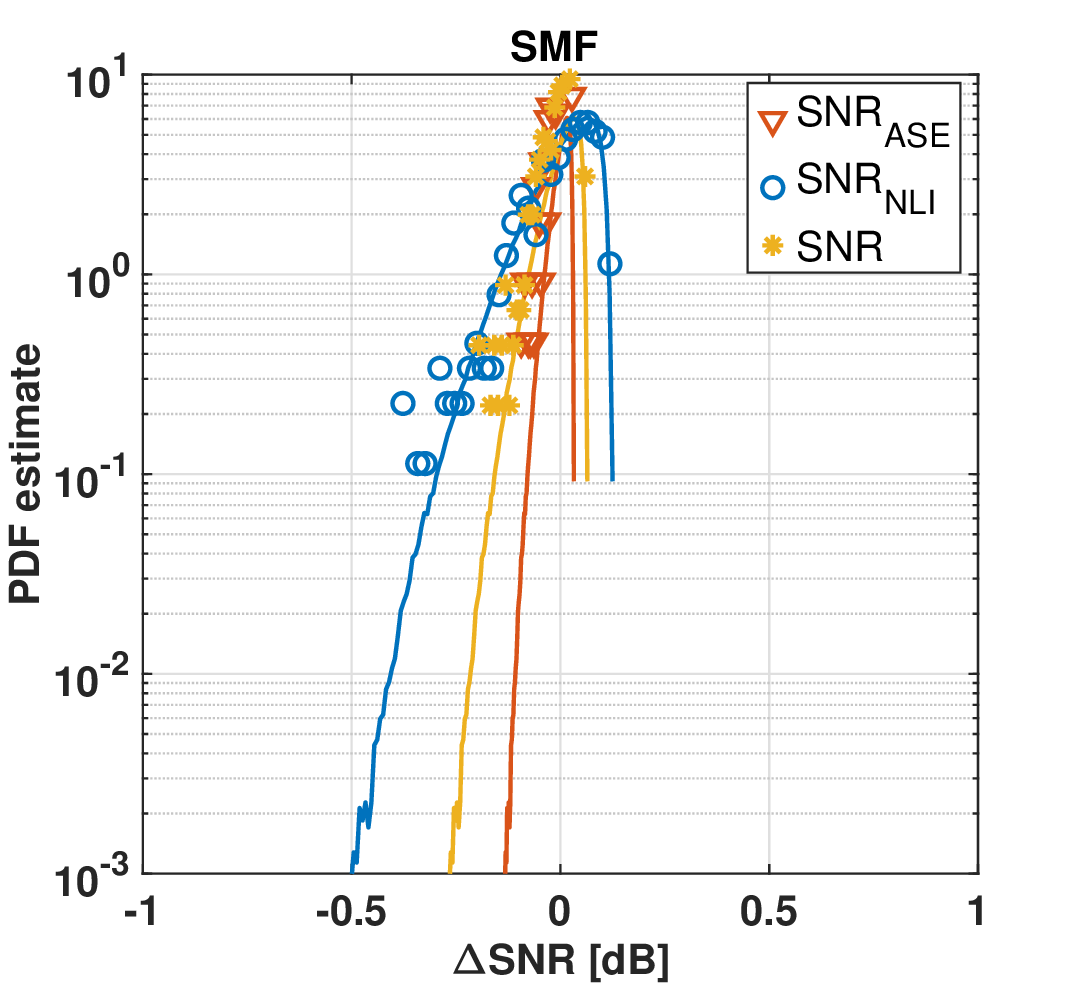}
\par\end{centering}
\begin{centering}
(b)
\par\end{centering}
\caption{PDF of $\Delta\mathrm{\text{SNR}}$, i.e., SNR deviation from its
mean, for the three cases under test in the absence of SMD. Top (a):
core 1 of a four-core CC-MCF, peak-to-peak MDL of each amplifier
equal to $1$ dB. Bottom (b): SMF, peak-to-peak PDL of each amplifier
equal to $0.5$ dB. Markers: PDF estimated from SSFM results. Solid
lines: PDL-GN model \cite{Serena20PDL} results. Channel power set
to maximize the SNR, in the absence of MDL, in both setups.\protect\label{fig:PDFs_noSMD}}
\end{figure}

As a preliminary investigation, we studied the SNR statistics in the
absence of SMD, which is more representative of SMF transmission rather
than CC-MCF. Figure \ref{fig:PDFs_noSMD}(a) shows, for the CC-MCF,
the PDF of the SNR in the three cases under test, i.e., linear, nonlinear,
and total SNR. We normalized the SNR to its mean value to ease the
comparison among the three distributions, namely: 
\begin{equation}
\Delta\mathrm{\text{SNR}\,[dB]\triangleq\mathrm{\text{SNR}\,[dB]-\left\langle \text{SNR}\,[dB]\right\rangle }}\,.\label{eq:=00005CDeltaSNR}
\end{equation}
Markers indicate results from SSFM simulations, while solid lines
represent semi-analytical estimations provided by the PDL-GN model
\cite{Serena20PDL}. The model and the SSFM results are remarkably
in agreement with each other. However, it is worth noting that we
simulated $500$ realizations with SSFM owing to the computational
complexity, while the semi-analytical model performed substantially
faster, allowing us to test $10^{6}$ seeds.

The different distributions of the linear and nonlinear SNR reported
in the figure indicate different interactions of MDL with the NLI
and ASE noise. In particular, the figure shows that the MDL causes
greater random variations in the linear SNR compared to the nonlinear
one. This observation is in contrast with the results reported in
the literature for SMF transmissions \cite{Rossi2019,Serena20PDL}.
To explore this discrepancy, we performed simulations with SMF fibers
as well. The lumped peak-to-peak MDL (here PDL) of each element was
set to $0.5$ dB. This choice ensures the same MDL-induced reduction
in average spectral efficiency per mode, in the linear regime, for
both the SMF and CC-MCF \cite{Andrusier2014}. The channel power was
decreased to its optimal value equal to $3$ dBm. Figure \ref{fig:PDFs_noSMD}(b)
shows, for the SMF, the PDFs of the linear, nonlinear, and total SNR
obtained with numerical simulations and the semi-analytical model. 

Comparing the distributions in Fig. \ref{fig:PDFs_noSMD}, it is
evident that the PDFs of the SNR in the CC-MCF link exhibit less skewness
compared to the SMF case, both in linear and nonlinear regime. A
qualitative explanation of this difference can be obtained by noting
that the PDFs of the ordered singular values of the MDL matrix $\mathbf{M}$
(the dotted curves in Fig. \ref{fig:MDL_histograms}) in turn exhibit
less skewness in the multi-core case compared to the SMF case.

\begin{figure}
\begin{centering}
\includegraphics[width=0.98\linewidth]{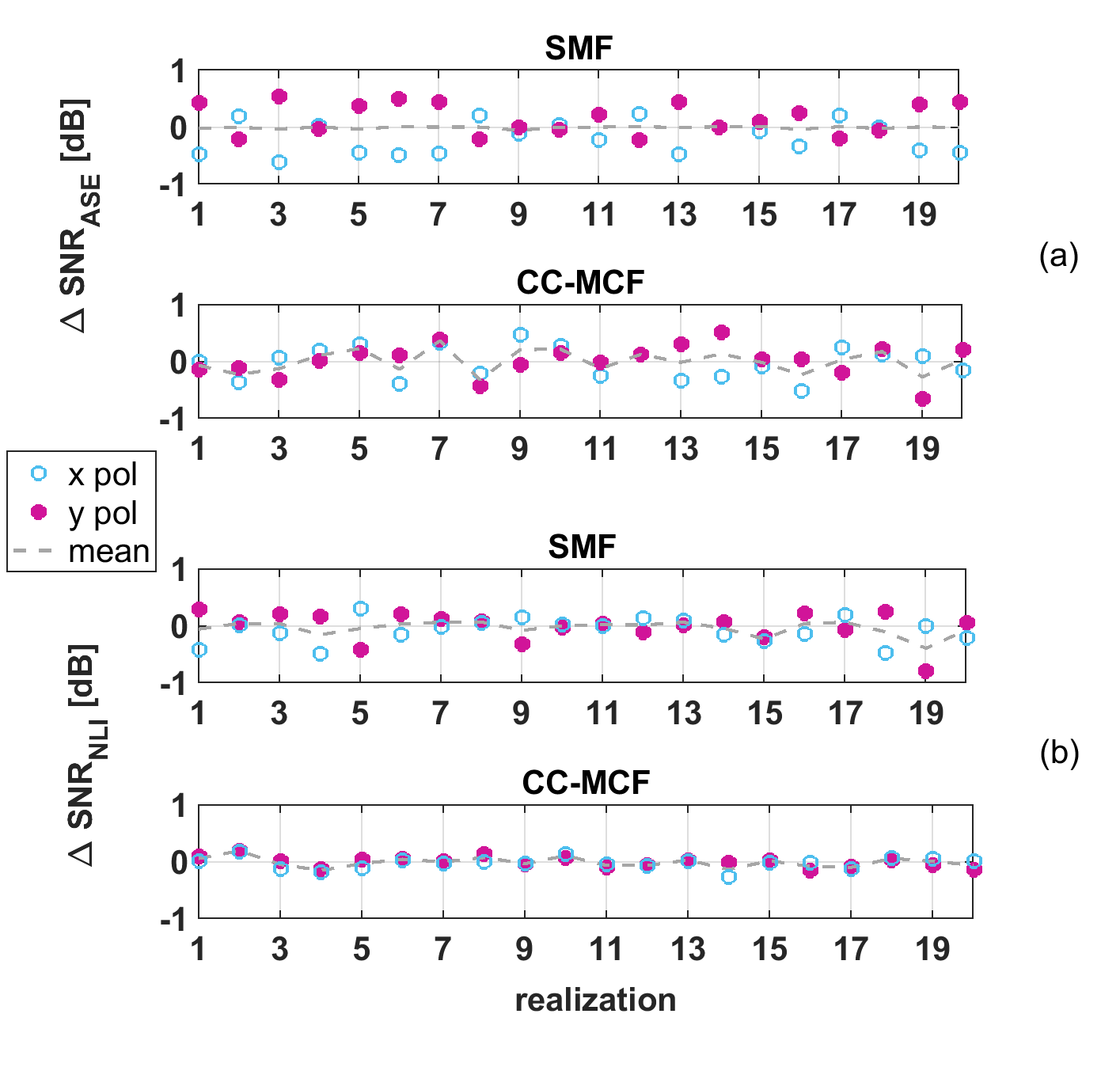}
\par\end{centering}
\caption{Markers: deviation from the mean of the per-polarization SNR of the
core under test in different random realizations, in the linear (a)
and nonlinear (b) case. Dashed line: average between polarizations.
Estimations obtained with the semi-analytical PDL-GN model. \protect\label{fig:subplot_SNRxy}}
\end{figure}
\begin{figure}[h]
\begin{centering}
\includegraphics[width=0.75\linewidth]{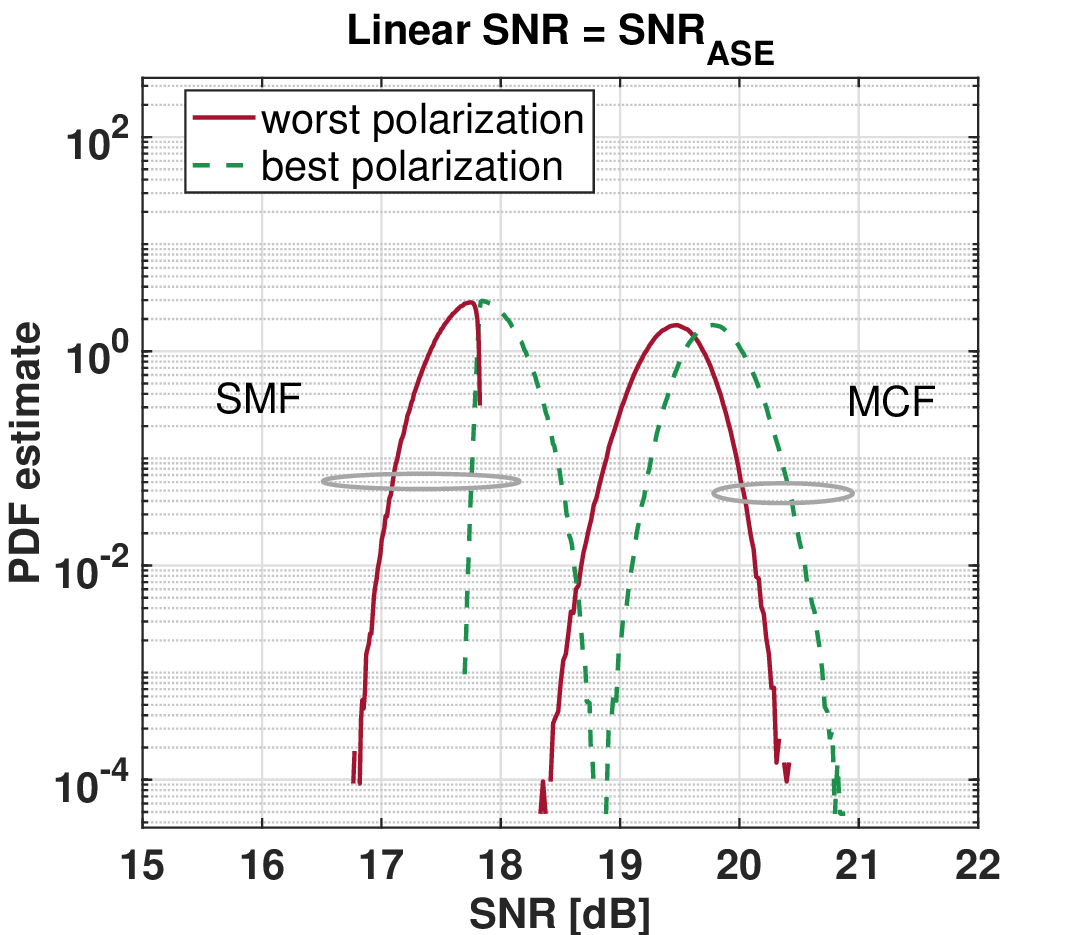}
\par\end{centering}
\begin{centering}
(a)
\par\end{centering}
\begin{centering}
\includegraphics[width=0.75\linewidth]{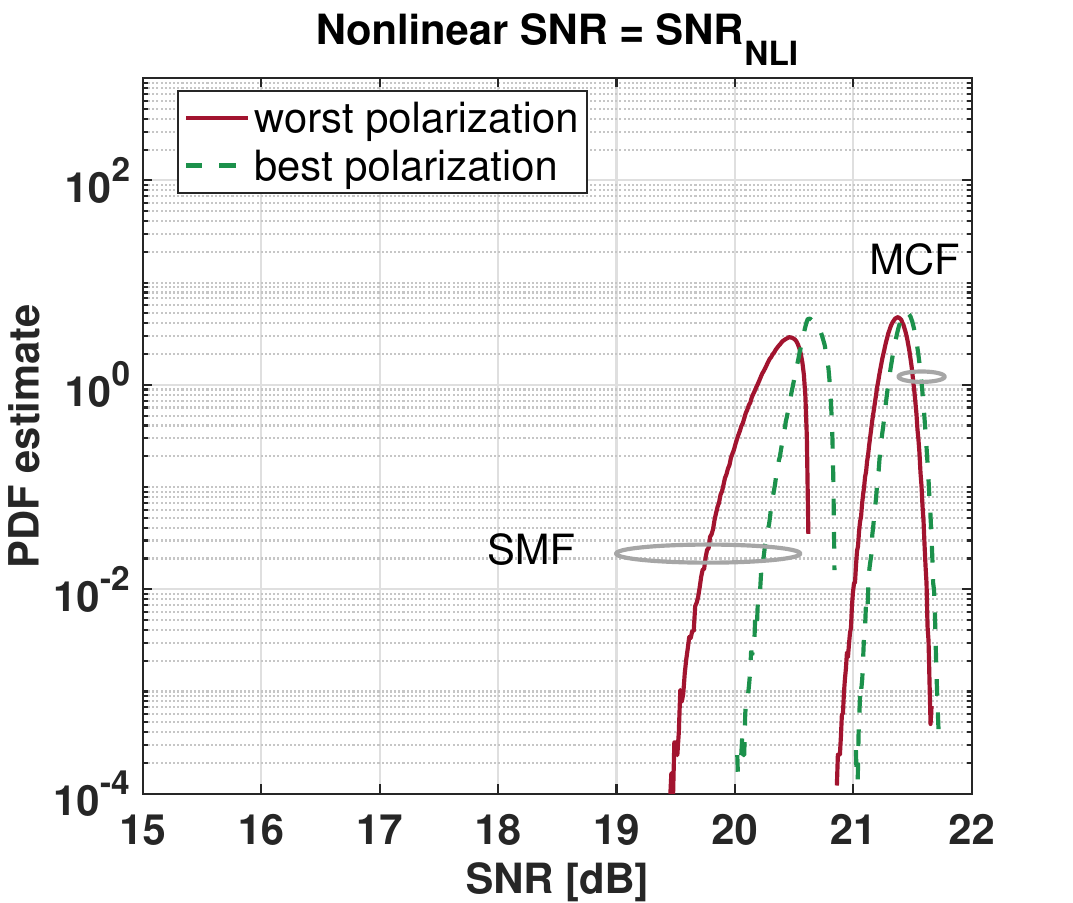}
\par\end{centering}
\begin{centering}
(b)
\par\end{centering}
\caption{PDF of linear (a) and nonlinear (b) SNR of the best and worst polarization
of the core under test, for the same scenario of Fig. \ref{fig:PDFs_noSMD},
obtained with the model in \cite{Serena20PDL}. Solid lines: four-core
CC-MCF. Dashed lines: SMF.\protect\label{fig:best_worst}}
\end{figure}

Another important difference emerging from the comparison between
the PDFs in Fig. \ref{fig:PDFs_noSMD} is the smaller standard deviation
in the SMF case, compared to the multi-core case. To gain insight
into this difference, we start by analyzing the linear SNR, that is
$\text{SNR}_{\text{ASE}}$. Figure \ref{fig:PDFs_noSMD} shows that
the SMF case has a smaller standard deviation than the CC-MCF by a
factor of $8.6$. The narrower PDF in the SMF case can be explained
considering the differences in the distributions of the MDL singular
values in Fig. \ref{fig:MDL_histograms}. While the single-mode PDL
increases the ASE power in one polarization, it depletes the other
following a symmetric bimodal distribution. As a confirmation of this
phenomenon, Fig. \ref{fig:subplot_SNRxy} (a) shows that the $\text{SNR}_{\text{ASE}}$
deviations from the mean exhibit antithetic behavior in the two polarizations,
leading to a fairly steady zero polarization-averaged value across
realizations (dashed line). As a consequence, the PDFs of the linear
SNR of the best- and worst-performing polarization exhibit antithetic
tails, as shown in Fig. \ref{fig:best_worst}(a). In contrast, in
the CC-MCF case, all the cores are mixed together, resulting in no
compensation between the polarizations within each core, as anticipated
in Sec. \ref{sec:MDLmatrix}. Consequently, the per-polarization linear
SNRs of the CUT may both decrease or increase, giving rise to a non-zero
average SNR deviation across the two polarizations in the same core,
as shown in the bottom panel of Fig. \ref{fig:subplot_SNRxy}(a).
As a result, the PDFs of the best and worst polarization in core 1
have comparable shapes.

In the nonlinear regime, the $\text{SNR}_{\text{NLI}}$ deviations
of the best and worst polarization in one core do not average to zero
in each realization, as shown in Fig. \ref{fig:subplot_SNRxy}(b).
This can be attributed to the different characteristics of the NLI
compared to the ASE noise. In particular, it is worth noting that
the self- and cross-phase modulation contributions (SPM and XPM, respectively)
to the NLI represent a scalar impairment which, despite the MDL-induced
unbalances, is common to all polarizations \cite{Rossi2019}. As a
result, the nonlinear SNR of the best and worst polarization exhibit
comparable, although shifted, distributions, as shown in Fig. \ref{fig:best_worst}(b).
In the SMF case, this leads to a wider PDF of the nonlinear SNR compared
to the linear case.    

\begin{figure}
\begin{centering}
\includegraphics[width=0.75\linewidth]{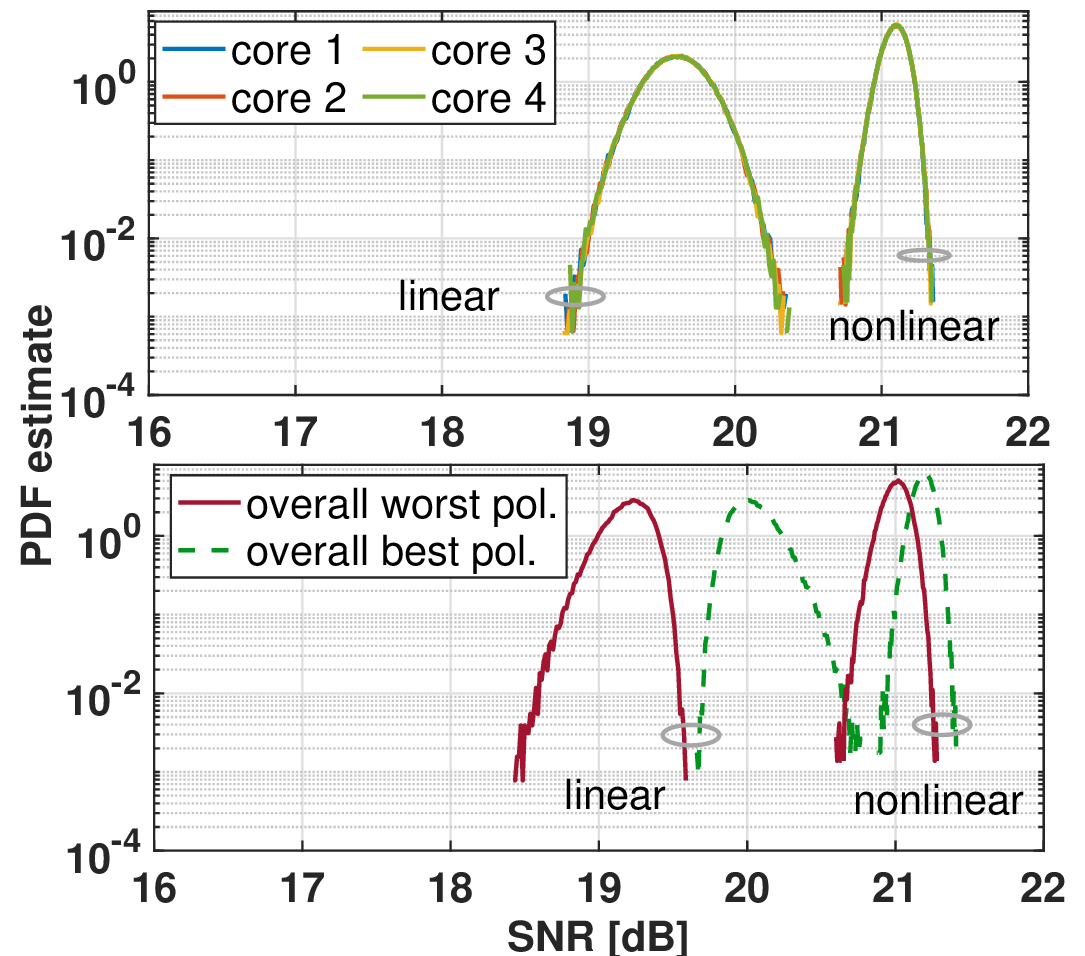}
\par\end{centering}
\caption{PDF of linear and nonlinear SNR of each core of the CC-MCF fiber (top),
and of the best and worst polarization among all the cores (bottom).\protect\label{fig:allcores_subplot}}
\end{figure}
\begin{figure*}[tbh]
\begin{centering}
\includegraphics[width=0.9\linewidth]{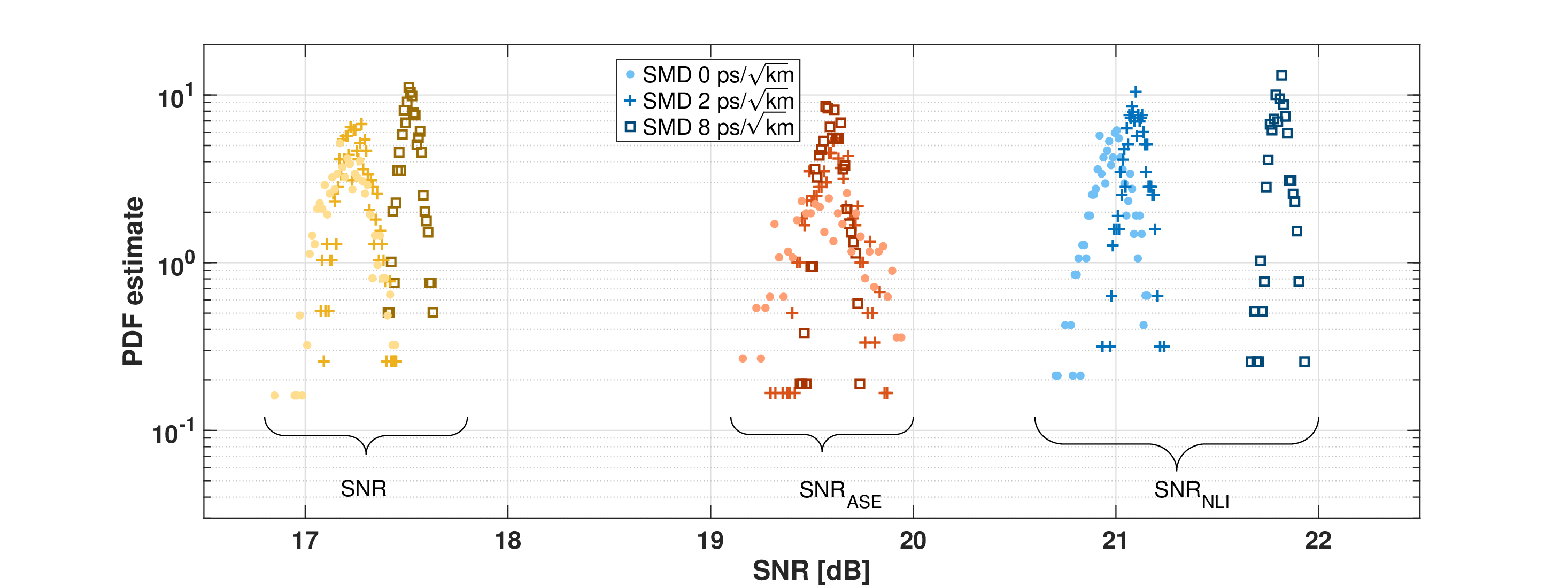}
\par\end{centering}
\caption{PDF of linear (center), nonlinear (right), and total (left) SNR estimated
from SSFM results collecting 500 channel realizations. $10\times100$
km link of MCF with four coupled cores, and peak-to-peak MDL of each
amplifier equal to $1$ dB. Fiber SMD coefficient equal to: $0$ (dots),
$2$ (crosses), and $8$ (squares) $\mathrm{\text{ps}/\sqrt{\text{km}}}$.
Transmitted WDM signal: five channels spaced $75$ GHz with symbol
rate equal to $64$ Gbaud and power $5$ dBm. \protect\label{fig:allPDFs} }
\end{figure*}

For completeness, Fig. \ref{fig:allcores_subplot} shows the PDF of
the linear and nonlinear SNR of each core of the CC-MCF. It can be
seen that, thanks to the strong linear mixing, all the cores are statistically
equivalent. The bottom panel of the figure shows that, when evaluated
across all the coupled cores, the PDFs of the worst- and best-performing
polarization exhibit antithetic tails in the linear regime, consistent
with the behavior observed in the SMF case, resulting in a vanishing
deviation of the polarization-averaged SNR when the average is calculated
across all cores

The results reported in this section were obtained neglecting mode
dispersion, whose effect is to make the system MDL frequency-dependent.
Such phenomenon, often referred to as frequency diversity \cite{Ho_frediversity_JLT11,Mello2020},
tends to average out MDL due to its random fluctuations across the
channel bandwidth.  Nevertheless, the extent of this effect on the
power of the NLI remains unclear. Therefore, the next section will
focus on estimating the SNR statistics using SSFM simulations that
include both SMD and MDL.

\subsection{SNR statistics with mode dispersion}

To investigate the combined effect of MDL and SMD on the SNR statistics,
we repeated the SSFM simulations including the SMD in the optical
fiber and full linear impairment compensation before detection, for
the same $10\times100$ km link.

The PDFs of the linear, nonlinear, and total SNR are reported in Fig.
\ref{fig:allPDFs}. The results obtained with an SMD coefficient of
$2$ $\mathrm{\text{ps}/\sqrt{\text{km}}}$ are represented with crosses,
while squares indicate the case of $8$ $\mathrm{\text{ps}/\sqrt{\text{km}}}$.
For comparison, we reported in the same figure the PDFs obtained in
the absence of SMD (dots). The figure shows that, in all the cases,
the SNR distributions remain Gaussian-like in the presence of mode
dispersion. Nevertheless, the differences among the displayed PDFs
indicate that the SMD affects differently the mean and standard deviation
of the SNR in the linear and nonlinear regime. This marks a significant
difference compared to the single-mode case, where the small value
of the PMD coefficient in typical SMF fibers has negligible impact
on the SNR distribution \cite{Serena20PDL}.  

It is important to note that our results were based on a zero-forcing
optical MIMO equalization rather than a least squares or minimum mean
square error equalizers \cite{Ospina24,Lucas_MMSE_JLT25,Gholamipourfard_JLT25}.
To assess the impact of this technique for the setup under test, we
performed additional simulations with a least-squares digital MIMO
approach finding no substantial difference with respect to the results
shown in Fig. \ref{fig:allPDFs}.

\begin{figure}
\begin{centering}
\includegraphics[width=0.75\linewidth]{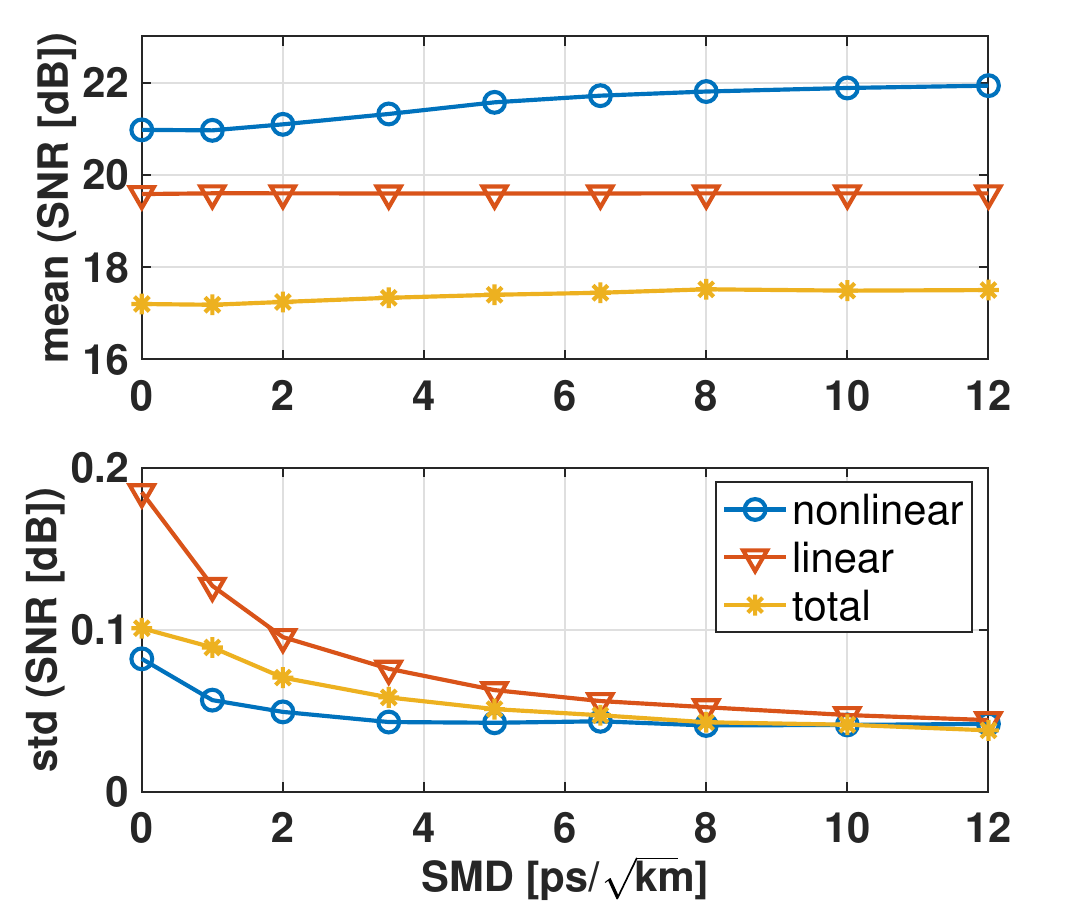}
\par\end{centering}
\caption{Standard deviation and mean value of the SNR as a function of the
SMD coefficient, estimated from SSFM results. Peak-to-peak MDL of
each amplifier equal to $1$ dB. \protect\label{fig:std_avg_SMD}}
\end{figure}

To better highlight the role of mode dispersion in setting the SNR
statistics, in Fig. \ref{fig:std_avg_SMD} we report the standard
deviation and mean value of the linear, nonlinear, and total SNR as
a function of the SMD coefficient, extending the range up to $12$
$\mathrm{\text{ps}/\sqrt{\text{km}}}$. Figure \ref{fig:std_avg_SMD}
shows that the mean SNR is affected by mode dispersion only in the
presence of fiber nonlinearity. Namely, the nonlinear SNR increases
thanks to the beneficial effect of the SMD in mitigating the accumulation
of the NLI along the distance. Such an observation is in agreement
with the findings of \cite{SerenaSDMGN,Lasagni23}, where the effect
of SMD on the NLI was investigated neglecting the MDL.

\begin{figure}[h]
\begin{centering}
\includegraphics[width=0.77\linewidth]{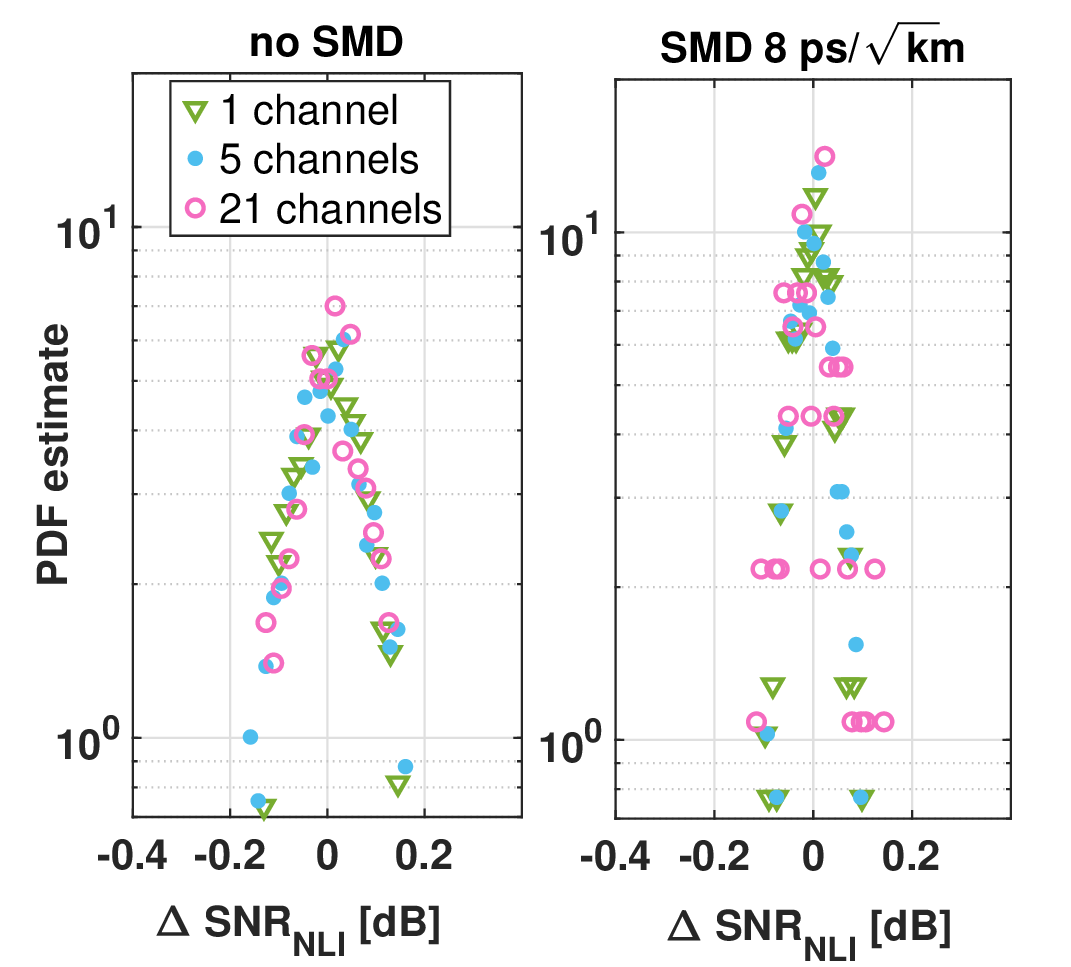}
\par\end{centering}
\caption{PDF of the nonlinear SNR deviation from its mean with SMD coefficient
$0$ $\text{ps}/\sqrt{\text{km}}$ (left) and $8$ $\mathrm{\text{ps}/\sqrt{\text{km}}}$
(right). Variable number of WDM channels: $1$ (triangles), $5$ (dots),
$21$ (circles), with channel power $5.5$ dBm, $5$ dBm, and $4.6$
dBm, respectively. Fixed symbol rate of $64$ Gbaud and channel spacing
$75$ GHz. Peak-to-peak MDL of each amplifier equal to $1$ dB.\protect\label{fig:PDFnonlinear_Nch}}
\end{figure}

\begin{figure}[tbh]
\begin{centering}
\includegraphics[width=0.72\linewidth]{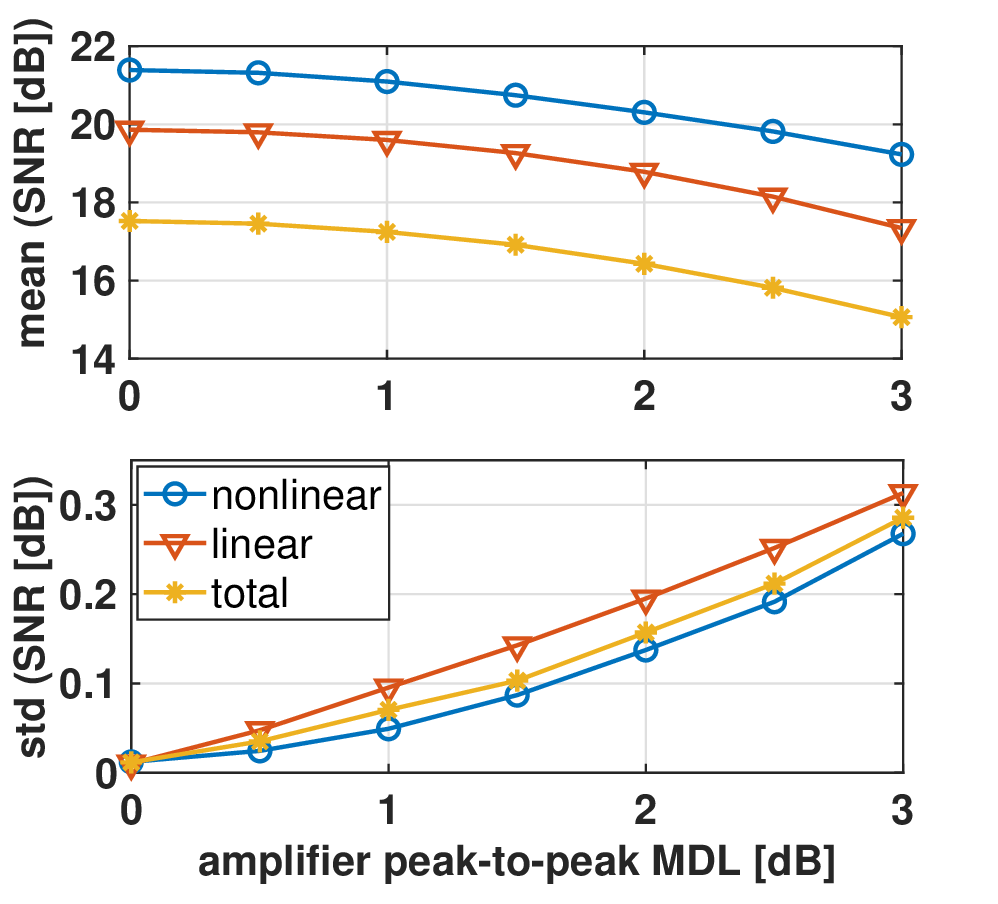}
\par\end{centering}
\caption{Standard deviation and mean value of the SNR as a function of the
peak-to-peak MDL of each link amplifier, estimated from SSFM results,
with fixed SMD coefficient of 2 $\mathrm{\text{ps}/\sqrt{\text{km}}}$.
\protect\label{fig:std_avg_MDL}}
\end{figure}
\begin{table*}
\centering{}\caption{Summary of Key Findings Across Analyzed Systems.\protect\label{tab:Summary-of-main}}
\setlength{\extrarowheight}{0.5cm}%
\begin{tabular}{ccccc}
\hline 
 & \begin{cellvarwidth}[m]
\centering
PDF of SNR\\
\end{cellvarwidth} & \begin{cellvarwidth}[m]
\centering
$\quad$$\quad$$\,\,$$\,\,$MDL-induced random fluctuations$\,\,$$\,\,$\\
\end{cellvarwidth} & \multicolumn{2}{c}{\begin{cellvarwidth}[m]
\centering
SNR behavior with mode dispersion\\
\end{cellvarwidth}}\tabularnewline
\hline 
\hline 
SMF & skewed & $\quad$$\text{\ensuremath{\mathrm{\text{std}}}(\text{\ensuremath{\mathrm{\text{NLI}}}}) \ensuremath{>} \ensuremath{\mathrm{\text{std}}}(\ensuremath{\mathrm{\text{ASE}}})}$ & \multicolumn{2}{c}{\begin{cellvarwidth}[t]
\centering
unaffected by practical values\\
\end{cellvarwidth}}\tabularnewline
\hline 
CC-MCF & symmetric & $\quad$$\mathrm{\text{std}}(\mathrm{\text{NLI}})<\mathrm{\text{std}}(\mathrm{\text{ASE}})$ & \begin{cellvarwidth}[m]
\centering
$\mathrm{\text{SNR}_{\text{NLI}}}$:\\
\vspace{0.1cm}$\quad\quad\quad$$\,$$\cdot$ decreased std\\
$\quad$$\quad\quad\quad$$\cdot$ increased mean\vspace{0.1cm}
\end{cellvarwidth} & \begin{cellvarwidth}[m]
\centering
$\mathrm{\text{SNR}_{\text{ASE}}}$:\\
\vspace{0.1cm}$\quad\quad\quad$$\cdot$ decreased std\\
$\quad\quad\quad\quad\,$$\cdot$ unaffected mean\vspace{0.1cm}
\end{cellvarwidth}\tabularnewline
\hline 
\end{tabular}
\end{table*}
Figure \ref{fig:std_avg_SMD} also shows that the SMD is effective
in reducing the standard deviation of the linear SNR. This behavior
is due to the random SMD-induced frequency fluctuations in the mode-dependent
gains/losses, which result in a beneficial averaging effect \cite{Mello2020,Ho2011,Antonelli_stokes}.
As a result, the random deviations of the ASE noise in core 1 are
reduced, leading to the shrinkage of the linear SNR PDF shown in Fig.
\ref{fig:allPDFs}. Such a frequency diversity phenomenon is still
present in the nonlinear noise, yet less evident. The results indicate
that the standard deviation of nonlinear SNR quickly saturates for
moderate values of the SMD coefficient. 

We observed that such behavior persists for different WDM bandwidths,
by comparing simulation results obtained with $1$, $5$, and $21$
channels. Figure \ref{fig:PDFnonlinear_Nch} shows the PDF of the
nonlinear SNR, normalized to its average value, obtained with an SMD
coefficient equal to $0$ $\text{ps}/\sqrt{\text{km}}$ (left) and
$8$ $\mathrm{\mathrm{\text{ps}/\sqrt{\text{km}}}}$ (right). For
a given SMD coefficient, the PDFs are comparable, indicating that
the inter-channel NLI (such as XPM and four-wave mixing) has little
effect on the PDF width.

We next investigated the impact of the per-amplifier MDL value on
the SNR statistics at a fixed SMD coefficient of $2$ $\mathrm{\text{ps}/\sqrt{\text{km}}}$.
Figure \ref{fig:std_avg_MDL} shows the standard deviation and the
mean value of the SNR in the three cases under investigation as a
function of the amplifiers' MDL. In particular, we note that the standard
deviation without MDL, i.e., caused by mode dispersion alone, is negligible
both in the linear and nonlinear regimes. Therefore, despite the random
nature of the SMD, the primary cause of SNR fluctuations can be attributed
to MDL. The results also show different growth rates of the standard
deviation of the two noises with the MDL magnitude. It is worth
noting that the range of MDL values tested here is in line with measured
peak-to-peak MDL values in the context of strongly coupled systems.
As an example, \cite{Mazur_ECOC21} reported an MDL below $3$ dB
at $1550$ nm for a seven-core CC-EDFA, \cite{Mazur_20} showed an
MDL below $2$ dB for a single span of a deployed four-core CC-MCF,
and the measurements in \cite{Ryf:19} for a four-core CC-MCF system
are consistent with a per-span MDL below $1$ dB. 

Figure \ref{fig:std_avg_MDL} also highlights the detrimental impact
of MDL on the mean SNR, showing a decrease of nearly $2.5$ dB in
the total mean SNR when each amplifier introduces $3$ dB of peak-to-peak
MDL. These results show that, while MDL is random, it eventually
manifests in the detected signal as a nearly deterministic degradation
of the SNR, particularly in the presence of SMD. These results, which
include NLI, are in line with the observations made in the linear
regime \cite{Antonelli_stokes,Ho2011,Lucas_JLT25}.

\section{Comments and Conclusions}

Mode-dependent loss is one of the most relevant propagation effects
that limit the capacity of SDM systems. Several experiments have shown
that different optical devices in the link may contribute significantly
to the accumulation of MDL. Understanding this phenomenon is therefore
essential for enhancing the capacity of SDM links. In this work, we
explored this problem from different perspectives, from an analysis
of the different implications of MDL on ASE and Kerr effect, to an
investigation on the role of SMD on the SNR statistics. We leveraged
the flexibility of a numerical analysis to isolate the Kerr contribution,
namely the NLI, and analyzed the physics of its interplay with MDL.

The main findings of this work are summarized in Tab. \ref{tab:Summary-of-main}.
Our results show that the fluctuations in the SNR induced by MDL are
primarily due to the interaction between MDL and ASE rather than MDL
and NLI, contrary to what has been observed in the literature for
polarization-dependent loss in SMFs. In addition, they revealed that
the SNR predictability is enhanced in the presence of SMD, thanks
to its frequency-averaging effect on MDL. Finally, we confirmed that
SMD, when equalized at the MIMO receiver, improves the SNR by mitigating
the Kerr effect even in the presence of MDL.

\end{document}